\documentclass[RNAAS]{aastex631}

\begin{document}

\title{\large Stellar Activity Cycles Grow Longer and Weaker Before Disappearing}

\author[0000-0003-4034-0416]{Travis S.~Metcalfe}
\affiliation{Center for Solar-Stellar Connections, WDRC, 9020 Brumm Trail, Golden, CO 80403, USA}

\begin{abstract}

In 2007, Erika B\"ohm-Vitense published a provocative figure suggesting that the solar 
rotation period and activity cycle made the Sun an outlier compared to the trends 
observed for stars in the Mount Wilson HK survey. A decade later, after the discovery of 
weakened magnetic braking (WMB), an evolutionary scenario was proposed that could account 
for the properties of the Sun if activity cycles grow longer and weaker in the WMB 
regime. Recent observations of the gradual onset of WMB suggest that the efficiency of 
the global stellar dynamo declines by at least two orders of magnitude as the stellar 
Rossby number approaches a critical point slightly above the solar value. A new sample of 
activity cycle data from the California Legacy Survey suggests that the Sun is not an 
outlier, and unambiguously confirms that activity cycles grow longer and weaker on 
stellar evolutionary timescales.

\end{abstract}

%%%%%%%%%%%%%%%%%%%%%%%%%%%%%%%%%%%%%%%%%%%%%%%%%%%%%%%%%%%%%%%%%%%%%%%%%% 

\section{1.~Astrophysical Context}

Nearly two decades ago, \cite{BohmVitense2007} published an analysis of stars from the 
Mount Wilson HK survey \citep{Baliunas1995} with measured rotation periods and the 
highest quality activity cycles. In the first figure of the paper she plotted the 
rotation periods against the cycle periods, revealing two distinct relationships between 
these quantities that she labeled the ``active'' and ``inactive'' sequences. In a 
provocative move she included the solar data point in the figure, illustrating 
dramatically that the 25.4~day sidereal rotation period and 11~yr activity cycle of the 
Sun made it an outlier---falling directly between the two stellar sequences.

Shortly after the discovery of weakened magnetic braking \citep[WMB;][]{vanSaders2016} in 
old field stars observed by the Kepler mission, \cite{Metcalfe2017} suggested a possible 
explanation for why the Sun appeared to be an outlier in the B\"ohm-Vitense plot. By 
color coding the data points to indicate the spectral type of the star, it became 
apparent that hotter stars only showed clear activity cycles at the shortest rotation 
periods, while cycles in solar analogs seemed to disappear at intermediate rotation 
periods and only cooler stars showed cycles at the longest rotation periods. Adding the 
measured rotation periods of ``flat-activity'' stars along the top of the figure, this 
pattern appeared to persist in stars without activity cycles. Finally, including data for 
several well-characterized solar analogs with precise asteroseismic ages (18\,Sco, 
$\alpha$\,Cen\,A, 16\,Cyg) suggested an evolutionary scenario in which activity cycles 
gradually grow longer and weaker in stars that have already entered the WMB regime. 
Consequently, the solar cycle might appear to be an outlier because it is slowly growing 
longer and weaker on stellar evolutionary timescales.

Observational constraints on the wind braking torque for 17 bright stars from the Mount 
Wilson HK survey have revealed the gradual onset of WMB as the stellar Rossby number (Ro) 
approaches a critical point slightly above the solar value \citep{Metcalfe2025}. The 
sources of the unexpected decline in wind braking torque can be seen directly in 
observations of the large-scale magnetic field strengths and the X-ray luminosities. The 
transition is abrupt in Ro, but it plays out slowly over the second half of main-sequence 
lifetimes as the efficiency of the global dynamo (and of magnetic braking) declines by at 
least two orders of magnitude. This leads to a pile-up of stars near the critical Ro, as 
rotation periods evolve very slowly under the influence of WMB \citep{vanSaders2019}. In 
the following section, I revisit the contemporaneous evolution of stellar activity cycles 
using a newly available data set.

\section{2.~Results}

\begin{figure*}[t]
 \centering\includegraphics[width=6.25in]{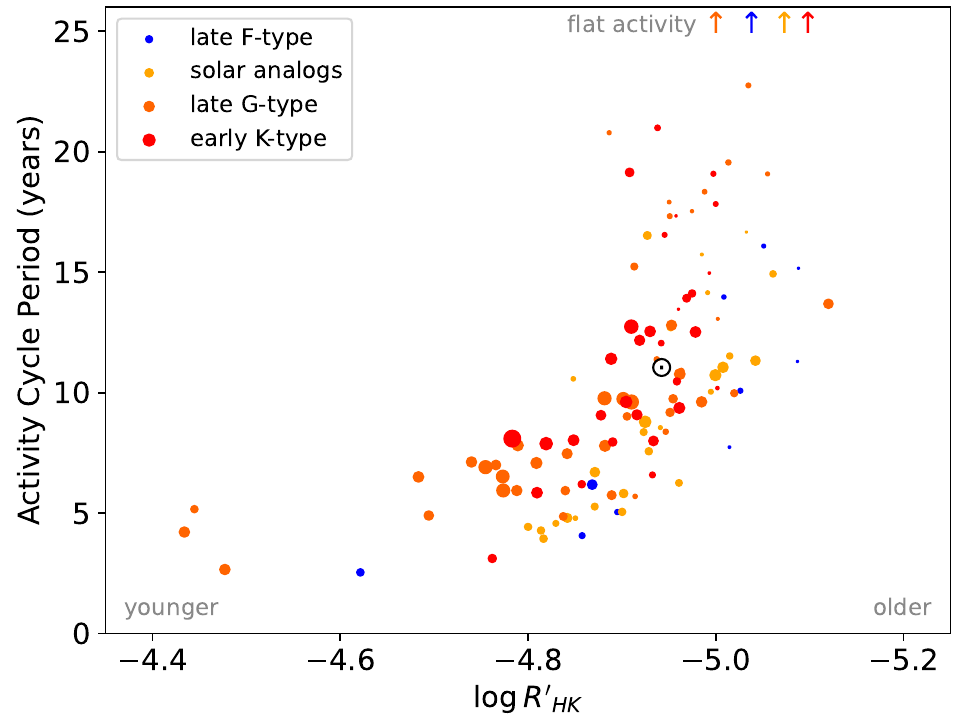}
 \caption{\sl Activity cycle period as a function of chromospheric activity level (a 
proxy for stellar age) for stars in the California Legacy Survey \citep{Isaacson2024, 
Isaacson2025}. Points are colored by spectral type as indicated in the legend, and the 
point size indicates the relative cycle amplitude. The activity levels of several 
flat-activity stars are shown with arrows along the top.\label{fig1}}
 \end{figure*}

A large sample of activity cycle data was recently published by the California Legacy 
Survey \citep[CLS;][]{Isaacson2024, Isaacson2025}. Rotation periods for the stars are not 
available, but the chromospheric activity level ($\log R'_{\rm HK}$) is strongly 
correlated with Ro \citep{Brandenburg1998} and unlike the rotation period it evolves 
continuously with stellar age \citep{LorenzoOliveira2018}. Motivated by these 
developments, I substitute $\log R'_{\rm HK}$ for rotation period in the B\"ohm-Vitense 
figure and I plot the data from CLS in Figure~\ref{fig1}. The point color reflects the 
spectral type as indicated in the legend, and the point size indicates the relative 
amplitude of the activity cycle. The activity levels of several flat-activity stars 
($\rho$\,CrB, 16\,Cyg, 61\,Vir, 31\,Aql) are shown with arrows along the top.

% FIGURE 1 ---------------------------------------------------------------   
%-------------------------------------------------------------------------               

In this updated representation the Sun is no longer an outlier, and the CLS data support 
the evolutionary scenario proposed by \cite{Metcalfe2017}. On the left side of 
Figure~\ref{fig1}, at activity levels well above the solar value \citep{Egeland2017}, 
cycle periods grow longer relatively slowly as stars evolve according to the 
\cite{Skumanich1972} relations. As stars approach the solar activity level (and the 
critical value of Ro) the cycle periods grow longer more quickly, and the cycle 
amplitudes decline toward the low activity levels where flat-activity stars are found. 
Dynamo simulations suggest that magnetic grand minima (like the Sun's Maunder minimum) 
may become more prolonged and more frequent during this transition \citep{Vashishth2023}, 
so the observed flat-activity stars might include a mixture of temporarily and 
permanently dormant activity cycles. The horizontal scatter in Figure~\ref{fig1} may be 
due to residual dependencies of $\log R'_{\rm HK}$ on the stellar mass, metallicity, and 
surface gravity.

In summary, the Sun appears to be doing what the other stars in the figure are 
doing---it's just that they are all evolving in a way that was unanticipated in 2007. 
There are now two distinct indications that WMB is driven by dynamo evolution: [1] the 
gradual onset of WMB follows the functional form predicted for a supercritical Hopf 
bifurcation \citep{Cameron2017}, and [2] the activity cycle properties evolve in a manner 
that might be expected from a gradual decline in the efficiency of the global stellar 
dynamo. If the solar dynamo represents an inefficient version of the global dynamos found 
in other stars, it means that spatially-resolved observations of the Sun can be used to 
inform our understanding of dynamo evolution more generally.

%%%%%%%%%%%%%%%%%%%%%%%%%%%%%%%%%%%%%%%%%%%%%%%%%%%%%%%%%%%%%%%%%%%%%%%%%% 
\begin{acknowledgments}
T.S.M.\ acknowledges support from NASA grant 80NSSC25K7563 and NSF grant AST-2507890. 
\end{acknowledgments}

%%%%%%%%%%%%%%%%%%%%%%%%%%%%%%%%%%%%%%%%%%%%%%%%%%%%%%%%%%%%%%%%%%%%%%%%%% 

\newpage

\section{Appendix}

The data behind Figure~\ref{fig1} are from H.~Isaacson (private communication), and 
portions were originally published in \cite[][Tables 2 \& 4]{Isaacson2024}. The 
spectroscopic properties were used to calculate B$-$V, which was subsequently 
combined with $S_{\rm med}$ to calculate $\log R'_{\rm HK}$ on the Mount Wilson 
scale. The full sample was restricted to stars with $4700 < T_{\rm eff} < 5900$~K, 
$\log g > 4.3$, and $0.7 < M_\star < 1.1\ M_\odot$. The color coding for spectral 
type was based on the stellar mass (early K-type: 0.7--0.8 $M_\odot$, late G-type: 
0.8--0.9 $M_\odot$, solar analogs: 0.9--1.0 $M_\odot$, late F-type: 1.0--1.1 
$M_\odot$).

% TABLE 1 ----------------------------------------------------------------
 \startlongtable
 \begin{deluxetable}{lccRcccCcr}
 \tablecaption{Data Behind the Figure (H.~Isaacson, private communication) \label{tab1}}
 \tablehead{ \colhead{HD} & \colhead{$S_{\rm med}$} & \colhead{$T_{\rm eff}$} & \colhead{[Fe/H]} & \colhead{$\log g$} & \colhead{$M_\star$} & \colhead{B$-$V} & \colhead{$\log R'_{\rm HK}$} & \colhead{$A_{\rm cyc}$} & \colhead{$P_{\rm cyc}$} }
 \startdata
1461    & 0.1590 & 5695.2 &  0.182 & 4.327 & 1.031 & 0.669 & -5.009 & 0.0027 & 13.97 \\
3651    & 0.1665 & 5205.9 &  0.168 & 4.505 & 0.888 & 0.829 & -5.020 & 0.0087 &  9.98 \\
3765    & 0.1951 & 5049.8 &  0.172 & 4.527 & 0.852 & 0.889 & -4.953 & 0.0193 & 12.80 \\
4256    & 0.2396 & 4921.8 &  0.256 & 4.545 & 0.848 & 0.955 & -4.902 & 0.0321 &  9.74 \\
4628    & 0.1927 & 4997.9 & -0.235 & 4.602 & 0.738 & 0.850 & -4.934 & 0.0175 &  8.00 \\
4747    & 0.2492 & 5331.9 & -0.166 & 4.557 & 0.832 & 0.736 & -4.694 & 0.0168 &  4.90 \\
4915    & 0.1887 & 5667.5 & -0.151 & 4.486 & 0.898 & 0.632 & -4.838 & 0.0106 &  4.86 \\
7924    & 0.2117 & 5171.1 & -0.123 & 4.555 & 0.802 & 0.799 & -4.842 & 0.0187 &  7.47 \\
8389    & 0.1746 & 5244.5 &  0.406 & 4.440 & 0.945 & 0.855 & -5.000 & 0.0240 & 10.73 \\
9407    & 0.1629 & 5622.3 &  0.029 & 4.366 & 0.929 & 0.669 & -4.985 & 0.0009 & 15.74 \\
9986    & 0.1762 & 5795.8 &  0.089 & 4.397 & 1.032 & 0.626 & -4.895 & 0.0043 &  5.04 \\
10476   & 0.1926 & 5215.3 &  0.004 & 4.516 & 0.842 & 0.801 & -4.905 & 0.0102 &  9.02 \\
10700   & 0.1674 & 5461.1 & -0.417 & 4.508 & 0.755 & 0.664 & -4.958 & 0.0005 & 17.34 \\
12051   & 0.1585 & 5472.0 &  0.230 & 4.407 & 0.967 & 0.746 & -5.033 & 0.0004 & 16.67 \\
14412   & 0.1905 & 5388.2 & -0.405 & 4.558 & 0.766 & 0.689 & -4.857 & 0.0097 &  6.20 \\
16160   & 0.2260 & 4813.5 & -0.036 & 4.584 & 0.752 & 0.954 & -4.930 & 0.0216 & 12.54 \\
18143   & 0.1661 & 4818.6 &  0.168 & 4.598 & 0.833 & 0.984 & -5.120 & 0.0168 & 13.69 \\
18803   & 0.1802 & 5623.5 &  0.128 & 4.412 & 0.984 & 0.683 & -4.900 & 0.0095 &  5.05 \\
20165   & 0.2055 & 5137.8 &  0.023 & 4.541 & 0.840 & 0.832 & -4.882 & 0.0236 &  7.80 \\
20619   & 0.1955 & 5717.8 & -0.172 & 4.488 & 0.910 & 0.614 & -4.800 & 0.0081 &  4.43 \\
22049   & 0.4593 & 5020.4 & -0.044 & 4.590 & 0.815 & 0.867 & -4.477 & 0.0208 &  2.66 \\
24238   & 0.1668 & 5202.3 & -0.357 & 4.557 & 0.722 & 0.758 & -4.993 & 0.0006 & 14.97 \\
24496   & 0.1799 & 5575.0 &  0.030 & 4.432 & 0.927 & 0.684 & -4.902 & 0.0131 &  5.81 \\
25665   & 0.2751 & 4962.6 &  0.021 & 4.567 & 0.805 & 0.900 & -4.773 & 0.0320 &  6.52 \\
26151   & 0.1850 & 5384.7 &  0.308 & 4.462 & 0.979 & 0.788 & -4.927 & 0.0111 & 16.52 \\
26965   & 0.1946 & 5246.0 & -0.198 & 4.520 & 0.771 & 0.762 & -4.878 & 0.0174 &  9.06 \\
28946   & 0.2174 & 5343.9 & -0.101 & 4.534 & 0.846 & 0.741 & -4.788 & 0.0200 &  5.94 \\
29883   & 0.1824 & 5044.8 & -0.094 & 4.556 & 0.777 & 0.851 & -4.969 & 0.0109 & 13.92 \\
32147   & 0.2598 & 4805.5 &  0.186 & 4.552 & 0.799 & 0.993 & -4.905 & 0.0259 &  9.61 \\
37008   & 0.1791 & 5208.4 & -0.322 & 4.555 & 0.733 & 0.760 & -4.938 & 0.0053 & 20.99 \\
38230   & 0.1602 & 5244.1 & -0.006 & 4.478 & 0.828 & 0.789 & -5.035 & 0.0037 & 22.75 \\
38858   & 0.1666 & 5736.4 & -0.172 & 4.451 & 0.900 & 0.609 & -4.941 & 0.0021 &  8.55 \\
42618   & 0.1589 & 5716.7 & -0.086 & 4.409 & 0.920 & 0.626 & -4.995 & 0.0038 & 10.04 \\
49674   & 0.1927 & 5628.4 &  0.300 & 4.426 & 1.064 & 0.708 & -4.858 & 0.0059 &  4.07 \\
51866   & 0.3156 & 4905.1 &  0.088 & 4.563 & 0.806 & 0.934 & -4.740 & 0.0206 &  7.12 \\
52711   & 0.1577 & 5814.8 & -0.119 & 4.333 & 0.929 & 0.593 & -4.991 & 0.0021 & 14.15 \\
62613   & 0.1837 & 5501.2 & -0.063 & 4.507 & 0.893 & 0.694 & -4.889 & 0.0145 &  5.75 \\
65277   & 0.2385 & 4770.7 & -0.059 & 4.593 & 0.737 & 0.969 & -4.919 & 0.0195 & 12.17 \\
68017   & 0.1756 & 5712.4 & -0.324 & 4.381 & 0.815 & 0.598 & -4.886 & 0.0024 & 20.79 \\
69830   & 0.1710 & 5420.6 &  0.012 & 4.490 & 0.890 & 0.731 & -4.963 & 0.0053 & 10.90 \\
72673   & 0.1719 & 5301.0 & -0.305 & 4.554 & 0.766 & 0.730 & -4.959 & 0.0092 & 10.47 \\
73667   & 0.1654 & 5208.1 & -0.409 & 4.564 & 0.713 & 0.750 & -4.997 & 0.0040 & 19.09 \\
75732   & 0.1689 & 5317.9 &  0.383 & 4.458 & 0.975 & 0.825 & -5.008 & 0.0206 & 11.05 \\
80606   & 0.1548 & 5565.9 &  0.348 & 4.402 & 1.047 & 0.735 & -5.051 & 0.0022 & 16.09 \\
86728   & 0.1471 & 5739.0 &  0.226 & 4.311 & 1.071 & 0.663 & -5.088 & 0.0004 & 15.17 \\
87359   & 0.2011 & 5668.2 &  0.070 & 4.436 & 0.982 & 0.661 & -4.800 & 0.0077 &  4.43 \\
87883   & 0.2679 & 4989.4 &  0.104 & 4.549 & 0.826 & 0.902 & -4.789 & 0.0231 &  7.81 \\
89269   & 0.1672 & 5635.7 & -0.138 & 4.424 & 0.871 & 0.643 & -4.951 & 0.0037 & 17.33 \\
90156   & 0.1629 & 5629.6 & -0.180 & 4.440 & 0.856 & 0.639 & -4.975 & 0.0012 & 17.53 \\
92719   & 0.1879 & 5774.6 & -0.103 & 4.425 & 0.937 & 0.607 & -4.830 & 0.0058 &  4.57 \\
97658   & 0.1864 & 5194.0 & -0.228 & 4.567 & 0.772 & 0.777 & -4.916 & 0.0194 &  9.08 \\
98281   & 0.1787 & 5452.7 & -0.161 & 4.489 & 0.830 & 0.697 & -4.913 & 0.0085 & 15.24 \\
99491   & 0.1976 & 5465.4 &  0.331 & 4.462 & 1.020 & 0.765 & -4.868 & 0.0174 &  6.18 \\
99492   & 0.2480 & 4936.5 &  0.278 & 4.539 & 0.856 & 0.953 & -4.882 & 0.0337 &  9.77 \\
100623  & 0.1914 & 5200.9 & -0.321 & 4.576 & 0.748 & 0.763 & -4.890 & 0.0129 &  7.95 \\
104067  & 0.3422 & 4939.0 &  0.087 & 4.564 & 0.818 & 0.920 & -4.683 & 0.0216 &  6.50 \\
104304  & 0.1592 & 5538.6 &  0.294 & 4.434 & 1.026 & 0.735 & -5.026 & 0.0040 & 10.08 \\
114783  & 0.1885 & 5126.7 &  0.151 & 4.526 & 0.867 & 0.856 & -4.951 & 0.0123 &  9.18 \\
116442  & 0.1638 & 5297.0 & -0.310 & 4.546 & 0.759 & 0.730 & -5.000 & 0.0036 & 17.83 \\
116443  & 0.1731 & 5145.0 & -0.289 & 4.577 & 0.742 & 0.787 & -4.975 & 0.0094 & 14.12 \\
122064  & 0.2473 & 4826.7 &  0.147 & 4.556 & 0.798 & 0.977 & -4.910 & 0.0353 & 12.74 \\
125455  & 0.1805 & 5155.5 & -0.088 & 4.549 & 0.806 & 0.810 & -4.955 & 0.0129 &  9.75 \\
126053  & 0.1642 & 5737.2 & -0.282 & 4.402 & 0.840 & 0.596 & -4.950 & 0.0018 & 17.91 \\
128311  & 0.5915 & 4931.9 &  0.134 & 4.574 & 0.837 & 0.930 & -4.434 & 0.0206 &  4.21 \\
130992  & 0.3227 & 4796.4 & -0.030 & 4.593 & 0.756 & 0.962 & -4.762 & 0.0126 &  3.12 \\
136352  & 0.1704 & 5746.5 & -0.254 & 4.358 & 0.846 & 0.597 & -4.914 & 0.0029 &  5.69 \\
136713  & 0.3104 & 4944.2 &  0.222 & 4.551 & 0.851 & 0.940 & -4.755 & 0.0340 &  6.91 \\
139323  & 0.2062 & 5114.1 &  0.349 & 4.503 & 0.910 & 0.894 & -4.925 & 0.0261 &  8.79 \\
140538A & 0.1977 & 5649.6 &  0.070 & 4.469 & 0.988 & 0.666 & -4.817 & 0.0100 &  3.94 \\
144287  & 0.1602 & 5480.4 & -0.007 & 4.404 & 0.873 & 0.709 & -5.013 & 0.0044 & 19.56 \\
144579  & 0.1664 & 5446.3 & -0.529 & 4.534 & 0.730 & 0.656 & -4.960 & 0.0004 & 13.46 \\
145675  & 0.1613 & 5314.9 &  0.405 & 4.427 & 0.969 & 0.829 & -5.042 & 0.0168 & 11.33 \\
145958A & 0.1751 & 5555.9 &  0.052 & 4.350 & 0.911 & 0.693 & -4.929 & 0.0095 &  7.57 \\
145958B & 0.1773 & 5521.2 &  0.058 & 4.369 & 0.904 & 0.705 & -4.923 & 0.0082 &  8.36 \\
146233  & 0.1655 & 5709.7 &  0.035 & 4.467 & 0.995 & 0.644 & -4.961 & 0.0079 &  6.25 \\
146362B & 0.2427 & 5886.4 & -0.018 & 4.440 & 1.026 & 0.587 & -4.621 & 0.0098 &  2.54 \\
149806  & 0.2029 & 5313.7 &  0.238 & 4.499 & 0.949 & 0.802 & -4.871 & 0.0168 &  6.70 \\
151541  & 0.1632 & 5376.8 & -0.114 & 4.496 & 0.831 & 0.728 & -5.002 & 0.0009 & 13.06 \\
154088  & 0.1550 & 5401.1 &  0.342 & 4.446 & 0.990 & 0.789 & -5.061 & 0.0071 & 14.93 \\
154345  & 0.2161 & 5490.6 & -0.070 & 4.505 & 0.886 & 0.697 & -4.766 & 0.0175 &  7.00 \\
155712  & 0.1892 & 4999.7 & -0.048 & 4.559 & 0.784 & 0.875 & -4.961 & 0.0229 &  9.37 \\
156279  & 0.1746 & 5458.2 &  0.192 & 4.450 & 0.960 & 0.745 & -4.952 & 0.0095 & 12.72 \\
156668  & 0.2331 & 4903.7 &  0.049 & 4.551 & 0.785 & 0.928 & -4.889 & 0.0247 & 11.41 \\
156985  & 0.2646 & 4863.9 & -0.082 & 4.585 & 0.749 & 0.925 & -4.819 & 0.0313 &  7.88 \\
158633  & 0.1736 & 5350.8 & -0.356 & 4.543 & 0.759 & 0.707 & -4.942 & 0.0051 & 12.06 \\
159062  & 0.1701 & 5491.9 & -0.272 & 4.468 & 0.791 & 0.671 & -4.946 & 0.0036 & 16.55 \\
166620  & 0.1706 & 5099.7 & -0.112 & 4.540 & 0.775 & 0.827 & -5.002 & 0.0015 & 10.19 \\
172051  & 0.1677 & 5617.3 & -0.207 & 4.484 & 0.858 & 0.640 & -4.947 & 0.0046 &  8.38 \\
176377  & 0.1805 & 5845.1 & -0.229 & 4.428 & 0.907 & 0.572 & -4.850 & 0.0030 &  4.79 \\
182488  & 0.1626 & 5416.6 &  0.199 & 4.465 & 0.954 & 0.760 & -5.015 & 0.0068 & 11.52 \\
185144  & 0.2066 & 5236.4 & -0.172 & 4.569 & 0.805 & 0.769 & -4.840 & 0.0129 &  5.93 \\
185414  & 0.1833 & 5804.4 & -0.095 & 4.412 & 0.949 & 0.599 & -4.848 & 0.0030 & 10.57 \\
187123  & 0.1565 & 5765.9 &  0.106 & 4.320 & 1.020 & 0.637 & -5.015 & 0.0007 &  7.74 \\
189733  & 0.5099 & 5012.5 &  0.041 & 4.571 & 0.828 & 0.883 & -4.445 & 0.0096 &  5.17 \\
190067  & 0.1743 & 5417.9 & -0.288 & 4.528 & 0.790 & 0.693 & -4.933 & 0.0058 &  6.59 \\
192310  & 0.1806 & 5181.1 &  0.081 & 4.494 & 0.844 & 0.825 & -4.962 & 0.0220 & 10.77 \\
196761  & 0.1721 & 5504.6 & -0.197 & 4.513 & 0.841 & 0.676 & -4.937 & 0.0049 & 11.38 \\
197076  & 0.1783 & 5833.0 & -0.065 & 4.429 & 0.979 & 0.595 & -4.871 & 0.0084 &  5.27 \\
202751  & 0.2079 & 4805.8 & -0.024 & 4.582 & 0.754 & 0.959 & -4.978 & 0.0216 & 12.52 \\
208313  & 0.2615 & 5027.7 &  0.010 & 4.564 & 0.818 & 0.872 & -4.774 & 0.0334 &  5.94 \\
212291  & 0.1975 & 5589.7 & -0.119 & 4.516 & 0.904 & 0.659 & -4.814 & 0.0094 &  4.28 \\
213042  & 0.3713 & 4730.6 &  0.199 & 4.567 & 0.792 & 1.028 & -4.783 & 0.0581 &  8.09 \\
215152  & 0.2486 & 4916.0 &  0.041 & 4.564 & 0.796 & 0.922 & -4.849 & 0.0223 &  8.03 \\
216520  & 0.1922 & 5160.0 & -0.124 & 4.545 & 0.791 & 0.803 & -4.908 & 0.0138 & 19.14 \\
217107  & 0.1487 & 5606.8 &  0.340 & 4.325 & 1.060 & 0.721 & -5.087 & 0.0004 & 11.30 \\
218566  & 0.2349 & 4942.9 &  0.298 & 4.529 & 0.855 & 0.953 & -4.910 & 0.0396 &  9.61 \\
218868  & 0.2001 & 5515.7 &  0.231 & 4.450 & 0.999 & 0.732 & -4.842 & 0.0148 &  4.80 \\
219538  & 0.2361 & 5094.0 &  0.008 & 4.550 & 0.827 & 0.846 & -4.809 & 0.0238 &  7.08 \\
219834B & 0.1778 & 5175.0 &  0.224 & 4.503 & 0.896 & 0.850 & -4.985 & 0.0204 &  9.62 \\
220339  & 0.2312 & 5052.2 & -0.213 & 4.587 & 0.751 & 0.831 & -4.810 & 0.0218 &  5.85 \\
221354  & 0.1564 & 5271.8 &  0.099 & 4.462 & 0.864 & 0.795 & -5.055 & 0.0023 & 19.08 \\
224619  & 0.1638 & 5494.8 & -0.091 & 4.460 & 0.859 & 0.693 & -4.988 & 0.0032 & 18.34 \\
 \enddata
 \end{deluxetable}
%-------------------------------------------------------------------------

\end{document}